\documentclass[runningheads,a4paper]{llncs}

\usepackage{amsmath,amsfonts,amssymb,textcomp,mathrsfs,dsfont}
\usepackage[hmargin=3.4cm,vmargin=3.6cm]{geometry}
\usepackage{graphicx}
\usepackage{subfig}
\usepackage{algorithm}
\usepackage{algorithmic}
\usepackage{url}

\newcommand{\set}[1]{\{{#1}\}}
\newcommand{\E}{\mathbb{E}}
\newcommand{\R}{\mathbb{R}}
\newcommand{\N}{\mathbb{N}}
\newcommand{\M}{{\mathcal M}}
\newcommand{\MM}{\M^{\delta,L}_{\mathfrak S}}
\newcommand{\A}{{\mathcal A}}
\newcommand{\tl}{\theta^{\ell}}
\newcommand{\V}    {\operatorname{Var}}

\newcommand{\Eps} {\varepsilon}
\newcommand{\ls}{{\scriptstyle \;\lesssim\;}}

\usepackage{xspace}
\newcommand\SHS{\mbox{ct-SHS}\xspace}
\begin{document}

\mainmatter  

\title{Multilevel Monte Carlo Method for Statistical\\ Model Checking of Hybrid Systems}

\titlerunning{MLMC Method for Statistical Model Checking of Hybrid Systems}

\author{Sadegh Esmaeil Zadeh Soudjani\inst{1} \and Rupak Majumdar\inst{1} \and Tigran Nagapetyan\inst{2}}

\authorrunning{S. Esmaeil Zadeh Soudjani \and R. Majumdar \and T. Nagapetyan}

\institute{Max Planck Institute for Software Systems, Kaiserslautern, Germany\\
\email{\{Sadegh,Rupak\}@mpi-sws.org}
\and
Department of Statistics, University of Oxford, United Kingdom\\
\email{Tigran.Nagapetyan@stats.ox.ac.uk}
}

\toctitle{Lecture Notes in Computer Science}
\tocauthor{Authors' Instructions}
\maketitle

\begin{abstract} 
We study statistical model checking of continuous-time stochastic hybrid systems.
The challenge in applying statistical model checking to these systems is that
one cannot simulate such systems exactly.
We employ the multilevel Monte Carlo method (MLMC) and work on 
a sequence of discrete-time stochastic processes whose executions approximate and converge 
weakly to that of the original continuous-time stochastic hybrid system with 
respect to satisfaction of the property of interest.
With focus on bounded-horizon reachability, we recast the model checking problem as the computation of the distribution of the exit time, 
which is in turn formulated as the expectation of an indicator function. 
This latter computation involves estimating discontinuous functionals, which reduces the bound on 
the convergence rate of the Monte Carlo algorithm. 
We propose a smoothing step with tunable precision and formally quantify the error of the 
MLMC approach in the mean-square sense, which is composed of smoothing error, bias, and variance.
We formulate a general adaptive algorithm which balances these error terms.
Finally, we describe an application of our technique
to verify a model of thermostatically controlled loads.

\end{abstract}

\section{Introduction}
\label{sec:intro}

Continuous-time stochastic hybrid systems (\SHS) are a natural model for cyber-physical
systems operating under uncertainty \cite{BujLyg04,CL06}.
A \SHS has a hybrid state space consisting of discrete modes and, for each mode, a set of continuous states (called the invariant). 
In each mode, the continuous state evolves according to a stochastic differential equation (SDE) in continuous time.
Transition from one discrete mode to another may be activated in two ways.
The continuous state may hit the boundary of the invariant and make a \emph{forced transition} according to a discrete stochastic
transition kernel.
Alternatively, the process may spontaneously change its discrete mode according to a continuous-time Markov chain 
whose rates depend on the hybrid state.

We consider quantitative analysis of temporal properties of \SHS \cite{BK08,BHHK03}.
The fundamental analysis problem, called \emph{probabilistic reachability}, consists 
in computing the probability that the state of a \SHS exits a given safe set within a given bounded time horizon.
Since analytic solutions are not available, there are two common approaches.
The first approach is \emph{numerical} model checking that relies on the exact or approximate computation of the measure 
of the executions satisfying the temporal property.
The second approach, called \emph{statistical} model checking, relies on finitely many sample executions of the system, 
and employs hypothesis testing to provide confidence intervals for the estimate of the probability.

Statistical model checking has proven to be computationally more efficient than numerical model 
checking as it only requires the system to be executable. 
Thus, it can be applied to larger classes of systems and of specifications \cite{SMC10}.
The main underlying assumption in all statistical model checking techniques is the ability 
to sample from the space of executions of the system.
Unfortunately, we cannot compute exact simulations for the general class of \SHS
due to the process evolution being continuous in both time and space.
In this paper, we describe a statistical model checking approach to \SHS using 
the \emph{multilevel Monte Carlo (MLMC)} method \cite{giles08,heinrich98}, 
which does not require exact executions of the system.

Our procedure works as follows.
First, we formulate the quantitative analysis problem as computing the distribution
of the first exit time of the system from the given safe set.
Then, we build a sequence of approximate models whose executions converge weakly (or in expectation) 
to the execution of the concrete system.
Although these approximate models can be used separately in the classical setting of statistical model checking
in order to compute estimates of the exit time, the MLMC method can take advantage of coupling between
approximate executions with different time resolutions to provide better convergence rates.

An important challenge in applying the MLMC technique to the quantitative analysis of \SHS is
that a discontinuous function is applied to the first exit time. 
While MLMC can be applied to discontinuous functions, the convergence rates we can guarantee
are poor.
We propose a smoothing step that replaces the discontinuous function with a continuous approximation
and show that the replacement decreases the overall computation cost.
Finally, we analyze the asymptotic computational cost of the MLMC approach for a given error bound.
We propose an adaptive algorithm which balances errors due to bias, variance, and smoothing,
and which tunes the hyperparameters of the algorithm on the fly.

We illustrate our technique on an example model of thermostatically controlled loads.

\smallskip
\noindent
\textbf{Related work.}
Formal definitions of various classes of \emph{continuous-time} probabilistic hybrid models are 
presented in \cite{PBLB03}, together with a comparison.
Over such models, \cite{Bujorianu2002} has formalized the notion of probabilistic reachability, 
\cite{PJP07} has proposed a computational technique based on convex optimization,
\cite{SADensity14} has provided discretization techniques with formal error bounds,
and \cite{FHT08} has developed an approach based on satisfiability modulo theory.
An alternative approach towards formal, finite approximations of continuous-time stochastic models is discussed 
in \cite{ZMMAL14} and extended in \cite{ZA14} to switching diffusions.
These approaches generally suffer from curse of dimensionality and are not applicable to large dimensional models.

For \emph{discrete-time} stochastic hybrid models
probabilistic reachability (and safety) has been fully characterized in \cite{APLS08} and 
computed via software tools \cite{SA13,SSoudjani} that use finite abstractions.
The methods can be extended to more general probabilistic temporal logics \cite{TMKA13}.
These techniques assume discrete-time dynamics and cannot be extended to \SHS.

An overview of statistical model checking techniques can be found in \cite{SMC10,Larsen14,K16}.
The paper \cite{DHKP17} employs statistical model checking for verifying unbounded temporal properties.
The paper \cite{Clarke2011} has discussed the use of importance sampling to address the issue of rare events in statistical verification of cyber-physical systems.
A distributed implementation of statistical model checking is proposed in \cite{Bulychev2012}
and a set-oriented method for statistical verification of dynamical systems is presented in \cite{WRWVD15}.

Employing multigrid ideas to reduce the computational complexity (in terms of expected number of arithmetic operations) of 
estimating an expected value using Monte Carlo path simulations is initially proposed in \cite{giles08} 
in the context of stochastic differential equations.
MLMC has a better asymptotic complexity and by its nature allows to build consecutive approximations, 
which can balance the bias and variance. 
The general paradigm with adequate modifications has shown significant gains in modeling jump-diffusion SDEs \cite{xg12} 
and in fault tolerance applications \cite{pa15}.
A more detailed overview of applications of MLMC can be found in \cite{giles15}.
The MLMC for estimating distribution functions is described in the recent paper \cite{Tigran17}, which is adapted to our setting.

\smallskip

The article is structured as follows. In Section \ref{sec:model}, we define the \SHS model and the probabilistic reachability 
problem.
In Sections~\ref{sec:Monte_Carlo} and~\ref{sec:MLMC}, we discuss the standard Monte Carlo technique
and the MLMC method, respectively, and compare their convergence rates. 
We then discuss two technical modifications: applying a smoothing operator to the discontinuous 
function of exit time (Section~\ref{s2.1}) and an 
adaptive MLMC algorithm for estimating the hyperparameters (Section~\ref{sec:algorithm}). 
In Section~\ref{sec:implementation}, we provide simulation results for an example.

\section{Model Definition}
\label{sec:model}

We study statistical model checking for the rich class of continuous-time stochastic hybrid systems (\SHS).

\subsection{Continuous-Time Stochastic Hybrid Systems}
\label{subsec:ct-SHS}

\begin{definition}
	\label{def:SHS_tuple}
	A \emph{continuous-time stochastic hybrid system} is a tuple $\mathcal H = \left(Q,\mathcal X, b, \sigma,x_0,r\right)$ 
	where the components are defined as follows.
	\begin{description}
		\item[States] 
		$Q$ is a countable set of discrete states (modes) and
		$\mathcal X: Q\rightarrow \mathcal P(\mathbb R^{n})$ maps each mode $q\in Q$ 
		to an open set $\mathcal{X}(q)\subseteq\mathbb R^n$, called the \emph{invariant} for the mode $q$.
		A state $(q,z)$ with $q\in Q$ and $z\in \mathcal{X}(q)$ is called a \emph{hybrid state}.
		The \emph{hybrid state space} $X$ is defined as 
		\begin{equation}
		\label{eq:hybrid_space}
		X = \set{(q, z)\mid q\in Q, z\in \mathcal{X}(q)}.
		\end{equation}
		We write $\partial Z$ for the boundary of a set $Z$ and define
		$\partial X := \set{(q,z)\mid q\in Q, z\in\partial \mathcal{X}(q)}$.
		
		\item[Evolution] 
		$b: X\rightarrow \mathbb R^n$ is a vector field	and $\sigma: X\rightarrow \mathbb R^{n\times m}$ is a matrix-valued function, with $n,m\in \mathbb N_0$,
		where $X$ is the hybrid space defined in \eqref{eq:hybrid_space}.
		For each $q\in Q$, define the following SDE:
		\begin{equation}
		\label{eq:SDEs}
		dz(t) = b(q,z(t))dt+\sigma(q,z(t))dW_t,
		\end{equation}
		where $(W_t,\,\,t\ge 0)$ is an $m$-dimensional standard Wiener process in a complete probability space.
		We assume functions $b(q,\cdot):\mathcal{X}(q)\rightarrow \mathbb R^n$ and 
		$\sigma(q,\cdot):\mathcal{X}(q)\rightarrow \mathbb R^{n\times m}$ are bounded and Lipschitz continuous for all $q\in Q$.
		The assumption ensures the existence and uniqueness of the solution of the SDEs in \eqref{eq:SDEs}.
		
		\item[Initial State] $x_0\in X$ is the initial state of the system; 
		\item[Transition Kernel]
		$r:\partial X\times Q\rightarrow[0,1]$ is a discrete stochastic kernel which governs the switching between the SDEs
		defined in \eqref{eq:SDEs}.
		That is, for all $q\in Q$, we assume $r(\cdot,q)$ is measurable and,
		for all $x\in\partial X$, the function $r(x,\cdot)$ is a discrete probability measure.
	\end{description}
\end{definition}

Intuitively, an execution of a \SHS starts in the initial state $x_0$,
and evolves according to the solution of the diffusion process \eqref{eq:SDEs}
for the current mode until it hits the boundary of the invariant of the current mode for the first time.
At this point, a new mode $q'$ is chosen according to the transition kernel $r$ and the execution
proceeds according to the solution of the diffusion process for $q'$, and so on.

We need the following definitions.
Let $z^q(t),\,\,q\in Q$ be the solution of diffusion process \eqref{eq:SDEs} starting from $z^q(0)\in \mathcal{X}(q)$.
Define $t^\ast(q)$ as the first exit time of $z^q(t)$ from the set $\mathcal{X}(q)$,
\begin{equation}
\label{eq:first_exit}
t^\ast(q) := \inf\{t\in\mathbb R_{>0}\cup\{\infty\},\,\text{ such that } z^q(t)\in\partial \mathcal{X}(q)\}.
\end{equation}
A stochastic hybrid process, describing the evolution of a \SHS, is obtained by the 
concatenation of diffusion processes $\{z^q(t),\,\,q\in Q\}$ together with a jumping mechanism given by 
a family of first exit times $t^\ast(q)$; we make this formal in Definition~\ref{eq:SHS_executions}.

\begin{definition}
	\label{eq:SHS_executions}
	A stochastic process $x(t) = (q(t),z(t))$ is called an \emph{execution} of \SHS
	$\mathcal H$ if there exists a sequence of stopping times $T_0 = 0<T_1<T_2<\ldots$ such that for all $k\in\mathbb N_0$:
	\begin{itemize}
		\item $x(0) = (q_0,z_0)\in X$ is the initial state of $\mathcal H$;
		\item for $t\in[T_k,T_{k+1})$, $q(t) = q(T_k)$ is constant and $z(t)$ is the solution of SDE
		\begin{equation*}
		dz(t) = b(q(T_k),z(t))dt+\sigma(q(T_k),z(t))dW_t,
		\end{equation*}
		where $W_t$ is the $m$-dimensional standard Wiener process;
		\item $T_{k+1} = T_k+t^\ast(q(T_k))$ where $t^\ast(q(T_k))$ is the first exit time from the mode $q(T_k)$ as defined in \eqref{eq:first_exit};
		\item The probability distribution of $q(T_{k+1})$ is governed by the discrete 
		kernel $r((q(T_k),z(T_{k+1}^-)),\cdot)$ and $z(T_{k+1}) = z(T_{k+1}^-)$, where $z(T_{k+1}^-) := \lim_{t\uparrow T_{k+1}}z(t)$.
	\end{itemize}
\end{definition}

\begin{remark}
	For simplicity of exposition, we have put the following restrictions on the \SHS
	model $\mathcal H$ in Definition \ref{def:SHS_tuple}.
	First, the model includes only forced jumps activated by reaching the boundaries of the invariant sets 
	$\partial \mathcal{X}(q),\,q\in Q$ and does not capture spontaneous jumps activated by Poisson processes. 
	Second, the continuous state $z(t)$ remains continuous at the switching times as declared in Definition \ref{eq:SHS_executions}.
	The approach of this paper is still applicable for \SHS models without these restrictions by modifying the 
	time discretization scheme presented in Section \ref{sec:Monte_Carlo}.
\end{remark}

\subsection{Example: Thermostatically Controlled Loads}
\label{sec:case_study}
Household appliances such as water boilers/heaters, air conditioners, and electric
heaters –-all referred to as \emph{thermostatically controlled loads (TCLs)}-– can store
energy due to their thermal mass. 
TCLs have been extensively studied \cite{SA14TCL,MESG13,MKC12}
for their role in energy management systems.
TCLs generally operate within a dead-band around a temperature set-point and
are naturally modeled using \SHS.
The temperature evolution in a cooling TCL can be characterized by the following SDE: 
\begin{equation}
\label{eq:tcl_dyn}
d\theta(t) = \frac{1}{CR}(\theta_a - q(t) R P_{rate}-\theta(t))dt+\sigma(q(t)) dW_t,
\end{equation}
where
$\theta_a$ is the ambient temperature,  
$P_{rate}$ is the energy transfer rate of the TCL,  
and $R$ and $C$ are the thermal resistance and capacitance, respectively.
The noise term $W_t$ in \eqref{eq:tcl_dyn} is a standard Wiener process.
The model of the TCL has two discrete modes.
When $q(t) = 0$, we say the TCL is in the OFF mode at time $t$,
and when $q(t) = 1$, we say it is in the ON mode.

The temperature of the cooling TCL is regulated by a control 
signal $q(t^+) = f(q(t),\theta(t))$ based on discrete switching as 
\begin{equation}
\label{eq:switch}
f(q,\theta)=
\left\{
\begin{array}{ll}
0, & \theta \le \theta_s - \delta_d/2 =: \theta_{-}\\ 
1,  & \theta \ge \theta_s + \delta_d/2 =: \theta_{+}\\
q, & \text{else,}
\end{array}
\right. 
\end{equation}
where $\theta_s$ denotes a temperature set-point and $\delta_d$ a dead-band.
Together, $\theta_s$ and $\delta_d$ characterize an operating temperature range.
The model can be described by the \SHS $\mathcal H_{TCL} = \left(Q,\mathcal X, b, \sigma,x_0,r\right)$, where
\begin{itemize}
	\item $Q = \{0,1\}$ with the invariants $\mathcal X(0) = (-\infty,\theta_+)$ and $\mathcal X(1) = (\theta_-,+\infty)$
	\item state space of the model $X = \{0\}\times (-\infty,\theta_+)\cup\{1\}\times (\theta_-,+\infty)$
	\item $b(q,\theta) = \frac{1}{CR}(\theta_a - q R P_{rate}-\theta)$ for all $(q,\theta)\in X$
	\item $\sigma(0,\theta) = \sigma(0),\sigma(1,\theta) = \sigma(1)$ for all $(q,\theta)\in X$
	\item $r(q^+\mid q,\theta)$ is the Kronecker delta with $q^+= f(q,\theta)$.
\end{itemize}

\subsection{Problem Definition}
\label{sec:safety}

For a given random variable defined on the executions of a \SHS, we study the problem of 
estimating its distribution function.

\begin{problem}
	\label{prob:distribution}
	Let $Y$ be a real-valued random variable defined on the 
	executions of \SHS $\mathcal H$. 
	Estimate $F_Y(s) := \mathbb{P}(Y\le s)$, the distribution of $Y$ for a given $s\in\mathbb R$.
\end{problem}

Consider a \SHS $\mathcal H$ with state space $X$,
a \emph{safe set} $A\subset X$, assumed to be measurable,
and a time interval $[0,s]\subset\mathbb R_{\ge 0}$.
The \emph{safety problem} asks to compute the probability that the 
executions of $\mathcal H$ will stay in $A$ during time interval $[0,s]$.
The safety problem is dual to the \emph{reachability problem} and has a fundamental role in model checking for \SHS.
By taking $Y$ in Problem~\ref{prob:distribution}
to be the first exit time of the system from $A$,
we reduce the safety problem to Problem~\ref{prob:distribution}.

\begin{problem}[Probabilistic Safety]
	\label{prob:safety}
	Compute the probability that an execution of the \SHS $\mathcal H$, 
	with initial condition $x_{0} \in X$, 
	remains within a measurable set $A$ during the bounded time horizon $[0,s]$:
	\begin{equation}
	\label{eq:safety}
	\mathbb P(\mathcal H\text{ is safe over }[0,s]) = \mathbb{P}(Y> s) = 1-F_Y(s)
	\end{equation}
	where $Y:= \min\{t\in\mathbb R_{\ge 0}\cup\{\infty\}\,|\, x(t)\notin A, x(0) = x_0\}$ and $F_Y(s) = \mathbb{P}(Y\le s)$.
\end{problem}

\begin{remark}
	The random variable $Y$ defined in Problem \ref{prob:safety} is in fact 
	the first exit time of the system $\mathcal H$ from the safe set $A$ 
	and its distribution can be represented as the expectation of an indicator functional:
	\begin{equation}
	\label{eq:expect_indicat}
	F_Y(s)=\E\left(\mathbf{1}_{(-\infty,s]} (Y)\right).
	\end{equation}
\end{remark}

\begin{problem}[Specification of interest for TCL] 
	\label{specinter}
	Although the switching mechanism \eqref{eq:switch} 
	is designed to keep the temperature inside the interval $[\theta_-,\theta_+]$, there is 
	still a chance that the temperature goes out of this interval due to the Wiener process $W_t$.
	Define a random variable $Y=\max\left\{\theta_t,\, t\in[0,s]\right\}$.
	We aim to estimate the probability $\mathbb{P}(Y\le\theta_++0.1\cdot \delta_d)$.
\end{problem}

Analytic solution of Problems~\ref{prob:distribution}-\ref{specinter} is infeasible for the class of \SHS. 
Numerical computation of the solution has been investigated for restrictive subclasses of \SHS~\cite{WSBP16,ABKCCL17}.
In this work, we propose an approximate computation technique with a confidence bound. 
Our technique based on MLMC substantially improves the computational complexity of the standard Monte Carlo method.
We first discuss standard Monte Carlo (SMC) method in Section \ref{sec:Monte_Carlo} and then present the MLMC method in Section \ref{sec:MLMC}.

\section{Standard Monte Carlo Method}
\label{sec:Monte_Carlo}

In order to compute the quantities of interest in Problems \ref{prob:distribution}-\ref{prob:safety}
we need to estimate \[\E P =  \E g\left(Y\right),\]
where $Y$ is a function of the execution of \SHS $\mathcal H$, $g:\mathbb R\rightarrow\mathbb R$ is the indicator function over the interval $(-\infty,s]$ and $P := g(Y)$ is a one-dimensional random variable.
The exact executions of $\mathcal H$ and thus exact samples of $Y$ are not available is general but it is possible to construct approximate executions and approximate samples that converge to the exact ones in a suitable sense.

Alg.~\ref{algo:Euler} presents a state update routine based on the \emph{Euler-Maruyama} method that can be used to 
construct approximate executions. 
Given the model $\mathcal H$ and the current approximate state $(q_k,z_k)$, this algorithm computes 
the approximate state $(q_{k+1},z_{k+1})$ for the next time step of size $\Delta$.
Equation \eqref{eq:Euler} in step \ref{step:Euler} of the algorithm is the Euler-Maruyama approximation of the SDE \eqref{eq:SDEs}.
If $z_{\textsf{aux}}$ is still inside the invariant of the current mode $\mathcal X(q_k)$, then the mode remains unchanged and $z_{\textsf{aux}}$ will be the next state (steps \ref{step:if}-\ref{step:same_mode}).
Otherwise, in steps \ref{step:project}-\ref{step:kernel} $z_{\textsf{aux}}$ is projected onto the boundary $\partial \mathcal{X}({q_k})$ of
the invariant
and the mode is updated according to the discrete kernel $r(q_k,z_{k+1})$.

\begin{algorithm}[t]
	\caption{State update $(q_{k+1},z_{k+1}) = \textsf{Update}(\mathcal H,q_k,z_k,\Delta,W_k)$}
	\begin{center}
		\begin{algorithmic}[1]
			\label{algo:Euler}
			\REQUIRE
			model $\mathcal H = \left(Q,\mathcal X, b, \sigma,x_0,r\right)$, current state $(q_k,z_k)$, time step $\Delta$, sampled noise $W_k$
			\STATE
			compute $z_{\textsf{aux}}$ according to the difference equation
			\label{step:Euler}
			\begin{equation}
			\label{eq:Euler}
			z_{\textsf{aux}} = z_{k} + b(q_k,z_k)\Delta+\sigma(q_k,z_k)\sqrt{\Delta}W_k
			\end{equation}
			\IF {$z_{\textsf{aux}}\in \mathcal X(q_k)$}
			\label{step:if}
			\STATE
			$z_{k+1}= z_{\textsf{aux}}$ and $q_{k+1} = q_k$
			\label{step:same_mode}
			\ELSE
			\STATE
			set $z_{k+1}$ to be the normal projection of $z_{\textsf{aux}}$ onto $\partial \mathcal{X}({q_k})$
			\label{step:project}
			\STATE
			select $q_{k+1}$ sampled from the distribution $r(q_k,z_{k+1})$
			\label{step:kernel}
			\ENDIF
			\ENSURE updated hybrid state $(q_{k+1},z_{k+1}) = \textsf{Update}(\mathcal H,q_k,z_k,\Delta,W_k)$
		\end{algorithmic}
	\end{center}
\end{algorithm}

Alg.~\ref{algo:dt-SHS} generates approximate executions of $\mathcal H$ and approximate samples of $Y$ 
using Alg.~\ref{algo:Euler}. 
The algorithm requires the model $\mathcal H$, the definition of $Y$ as a function of the execution of of $\mathcal H$, 
and the time interval $[0,s]$. The output of the algorithm $\theta^\ell$ is an approximate sample of random variable $Y$.
In steps \ref{step:ell}-\ref{step:Delta}
the number of time steps $n$ is selected and the discretization time step $\Delta$ is computed. In order to highlight the dependency of the algorithm to the parameter $n$, we have opted to use $\ell$ in the representation $n = \kappa 2^\ell$ as the superscript of the variables. We call $\ell$ the \emph{level} of approximation which is nicely connected to the MLMC terminology discussed in Section \ref{sec:MLMC}.

\begin{algorithm}[ht]
	\caption{Approximate sampling of random variable $Y$}
	\begin{center}
		\begin{algorithmic}[1]
			\label{algo:dt-SHS}
			\REQUIRE 
			model $\mathcal H = \left(Q,\mathcal X, b, \sigma,x_0,r\right)$, $Y$ a function of execution of $\mathcal H$, time interval $[0,s]$
			\STATE
			\label{step:ell}
			select the number of time steps $n$ and set $\kappa\ge 1,\ell\ge 0$ such that
			$n = \kappa 2^\ell$
			\STATE \label{step:Delta}
			compute the time step $\Delta := s/n$ and set $k:=0$
			\STATE \label{step:init}
			set the initial hybrid state $x^\ell_0 : = (q_0^\ell,z_0^\ell)$ according to $x_0 = (q_0,z_0)\in X$
			\WHILE {$k<n$}
			\STATE \label{step:sample}
			sample $W_k^\ell$ from the standard $m$-dimensional normal distribution
			\STATE \label{step:update}
			update the hybrid state $(q_{k+1}^\ell,z_{k+1}^\ell) = \textsf{Update}(\mathcal H,q_k^\ell,z_k^\ell,\Delta,W_k^\ell)$ using Alg.~\ref{algo:Euler}
			\STATE $ k = k+1$
			\ENDWHILE
			\STATE \label{step:interp}
			define for all $t\ge 0$, $z^\ell(t) = \sum\limits_{k=0}^{n}z_k^\ell 1_{[k\Delta,(k+1)\Delta)}(t)$ and $q^\ell(t) = \sum\limits_{k=0}^{n}q_k^\ell 1_{[k\Delta,(k+1)\Delta)}(t)$
			\STATE
			\label{step:sample_Y}
			compute $\theta^\ell$ by applying the definition of $Y$ to $(q^\ell(\cdot),z^\ell(\cdot))$
			\ENSURE
			$\theta^\ell$ as approximate sample of $Y$
		\end{algorithmic}
	\end{center}
\end{algorithm}

Alg.~\ref{algo:dt-SHS} initializes the approximate execution in step \ref{step:init} as $x^\ell_0 : = (q_0^\ell,z_0^\ell)$ according to $x_0$ the initial state of $\mathcal H$.
Then the algorithm iteratively computes the next approximate state $(q_{k+1}^\ell,z_{k+1}^\ell)$ by sampling from the $m$-dimensional standard normal distribution in step \ref{step:sample} and applying Alg.~\ref{algo:Euler} to $(\mathcal H, q_k^\ell,z_k^\ell,\Delta,W_k^\ell)$ in step \ref{step:update}.
Finally, step \ref{step:interp} constructs the continuous-time approximate execution $(q^\ell(\cdot),z^\ell(\cdot))$ as the piecewise constant version of the discrete execution $(q_k^\ell,z_k^\ell)$,
which enables the computation of $\theta^\ell$ by applying the definition of $Y$ to $(q^\ell(\cdot),z^\ell(\cdot))$ (step \ref{step:sample_Y}).

Alg.~\ref{algo:dt-SHS} is parameterized by $\ell$. 
Due to the nature of the Euler-Maruyama method in \eqref{eq:Euler}, we expect that the approximate samples $\theta^\ell$ 
converge to $Y$ as $\ell\rightarrow\infty$ in a suitable way. In fact, it is an unbiased estimator in the limit:
$\lim_{\ell\to\infty} \E g\left(\tl\right) = \E g\left(Y\right).$
The idea behind standard Monte Carlo (SMC) method is to use the empirical mean of $g\left(\tl\right)$ as an approximation of $\E g\left(Y\right)$.
The SMC estimator has the form
\begin{equation}
\label{eq:smc_est}
\hat{P} = \cfrac1N\sum\limits_{i=1}^N g\left(\tl_i\right),
\end{equation}
which is based on $N$ replications of $\tl$. 
The replications $\{\theta_i^\ell,i=1,\ldots,N\}$ can be generated by running Alg.~\ref{algo:dt-SHS} (with a fixed $\ell$)
$N$ times, or running any other algorithm that generates such samples (cf. Alg.~\ref{algo:dt-SHS:mlmc} in Section~\ref{sec:MLMC}).
The SMC method is summarized in Alg.~\ref{algo:SMC2}, which approximates $\E g(Y)$ based on a general sampling algorithm~$\A_\ell$.
Note that Alg.~\ref{algo:SMC2} can be used for estimating $\E g(Y)$ not only with $g(\cdot)$ being the indicator function but also any other functional that can be deterministically evaluated using the executions over the time interval $[0,s]$. 

Owing to the randomized nature of algorithm~$\A_\ell$ embedded in Alg.~\ref{algo:SMC2}, 
we quantify the quality of its outcome using \emph{mean squared error}:\footnote{
	We slightly abuse the notation and indicate by $MSE(\A_\ell)$ the mean square error of Alg.~\ref{algo:SMC2} with the embedded sampling algorithm~$\A_\ell$.}
\begin{equation}
\label{eq:mse}
\mathit{MSE}(\A_\ell) \equiv \E\left[\left(\hat{P} - \E P\right)^2\right] = 
\E\left[\left(\hat{P} - \E \hat{P}\right)^2\right] + \left[\E\hat{P} - \E P\right]^2.
\end{equation}
The mean square error $MSE(\A_\ell)$ is decomposed into two parts: 
\emph{Monte Carlo variance}
and
\emph{squared bias error}.
The latter is a systematic error arising from the fact that we might not sample our random variable exactly, but rather use a suitable approximation, 
while the former error comes from the randomized nature of the Monte Carlo algorithm.
The Monte Carlo variance (first term in \eqref{eq:mse}) is proportional to $N^{-1}$ as
$$\V\hat{P} = \V\left(\frac1N\sum\limits_{i=1}^N g(\tl_i)\right)=\frac{1}{N^2}\V\left(\sum\limits_{i=1}^N g(\tl_i)\right)=\frac1N\V\left(g(\tl)\right).$$
The cost of Alg.~\ref{algo:SMC2} is typically taken to be the expected runtime in order to achieve a prescribed accuracy $\mathit{MSE}(\A_\ell)\le \varepsilon$.
A more convenient approach for theoretical comparison between different methods is to consider the cost associated to sampling algorithm~$\A_\ell$,
$$C_\ell\left(\A_\ell\right) := \E\left[\text{\#operations and random number generations to calculate } g(\theta^\ell)\right],$$
which facilitates the definition of convergence rate of the algorithm.

\begin{definition}
	We say that Alg.~\ref{algo:SMC2} based on sampling algorithm~$\A_\ell$ 
	converges with rate $\gamma >0$ if $\lim\limits_{\ell\to \infty} \sqrt{MSE \left(\A_\ell\right)} = 0$ and if
	there exist constants $c > 0,\,\eta\ge0$ such that
	\begin{equation}
	\label{eq:rate}
	C_\ell \left(\A_\ell\right) \leq c \cdot \left(\sqrt{MSE \left(\A_\ell\right)}\right)^{-\gamma} \cdot
	\left( - \log \sqrt{MSE \left(\A_\ell\right)}\right)^{\eta}.
	\end{equation}
\end{definition}

\begin{remark}
	The definition of convergence rate in \eqref{eq:rate} indicates that for a desired accuracy $MSE \left(\A_\ell\right)\le\varepsilon$ smaller convergence rate $\gamma$ implies lower computational cost $C_\ell \left(\A_\ell\right)$.
\end{remark}

\begin{algorithm}[t]
	\caption{Standard Monte Carlo method to estimate $\E g(Y)$
	}
	\begin{center}
		\begin{algorithmic}[1]
			\label{algo:SMC2}
			\REQUIRE 
			Sampling algorithm~$\A_\ell$, 
			number of samples $N$, functional $g(\cdot)$
			\STATE $i:=1$
			\WHILE {$i<N$}
			\STATE
			sample $\theta_i^\ell$ using algorithm~$\A_\ell$ (for example Alg.~\ref{algo:dt-SHS} or Alg.~\ref{algo:dt-SHS:mlmc})
			\STATE evaluate $g(\theta_i^\ell)$
			\STATE $i = i+1$
			\ENDWHILE
			\ENSURE
			$\hat P = \frac1N\sum_{i=1}^Ng(\theta_i^\ell)$ as approximate estimate of $\E g(Y)$
		\end{algorithmic}
	\end{center}
\end{algorithm}

The following theorem presents the convergence rate of the SMC method presented in Alg.~\ref{algo:SMC2}.

\begin{theorem}
	\label{thm:SMC}
	Let $\theta^\ell$ denote the numerical approximation of the random variable $Y$ according to an algorithm~$\A_\ell$. Assume there exist positive constants $\alpha, \zeta, c_1, c_2$ such that for all $\ell\in\N_0$
	\begin{equation}
	\label{ass:SMC}
	\left| \mathbb{E}[g(\theta^\ell) \!-\! g(Y)] \right|\ \leq\ c_1 2^{-\alpha\cdot\ell},\,\quad
	\mathbb{E}[C_\ell]\ \leq\ c_2\, 2^{\zeta\cdot\ell},\quad\text{ and }
	\V g(\theta^\ell)\ <\ \infty.
	\end{equation}
	Then the standard Monte Carlo method of Alg.~\ref{algo:SMC2} based on sampling algorithm~$\A_\ell$ converges with rate $\gamma=2+\cfrac{\zeta}{\alpha}$.
\end{theorem}

\begin{remark}
	Recall the role of $\ell$ in step \ref{step:Delta} of Alg.~\ref{algo:dt-SHS}. Increasing $\ell$ results in an exponential increase in the number of time steps thus also in the number of samples. Therefore we have assumed in \eqref{ass:SMC} an exponential bound on the increased cost and an exponential bound in the decreased bias as a function of $\ell$. 
\end{remark}

\noindent
\textbf{Application to the TCL Case Study.}
We construct the approximate discrete-time executions as 
\begin{equation}
\theta^\ell_{k+1} = \frac{1}{CR}(\theta_a - q^\ell_k R P_{rate}-\theta^\ell_k)\Delta+\sigma(q^\ell_k)\cdot\sqrt{\Delta}\cdot W^{\ell}_k,
\label{euler:numer}
\end{equation}
where $W^{\ell}_k$ is the sample from the standard normal distribution, $\Delta=s/n$, $n = \kappa2^{\ell}$,
and the discrete mode at any level $\ell$ is defined as $q^\ell_{k+1} := f(q^\ell_k,\theta_k^\ell)$ with $f(\cdot)$ defined in \eqref{eq:switch}.
This discrete-time updating is slightly different from the $\textsf{Update}$ function of Alg.~\ref{algo:Euler}, which can be interpreted as follows. Instead of continuous updating of mode, the control signal acts as a digital controller and updates the mode only at the discrete time steps. It is clear, that the cost of simulating one execution of \eqref{euler:numer} is proportional to the number of the discretization steps, thus setting the parameter $\zeta=1$ in Theorem \ref{thm:SMC}.

\medskip

The values of constants $\alpha, \zeta, c_1, c_2$ in Theorem \ref{thm:SMC} depend on the regularity of the functional $g$, sampling algorithm~$\A_\ell$ and other parameters. In the next section we propose to use MLMC method that improves the convergence rate and substantially reduces the computational complexity of the estimation. We discuss a smoothing in Section \ref{s2.1} that replaces the indicator function $g(\cdot)$ with a smoothed function and discuss its effect on the algorithm's error.

\section{Multilevel Monte Carlo Method}
\label{sec:MLMC}

The multilevel Monte Carlo method (MLMC) relies on the simple observation of telescoping sum for expectation:
\begin{equation}
\label{eq:linearity}
\E g\left(\theta^{L}\right) = \E g\left(\theta^{0}\right) + \sum\limits_{l=1}^L \E\left[ g\left(\tl\right) - g\left(\theta^{\ell-1}\right)\right].
\end{equation}
where $\theta^0$ and $\theta^L$ correspond respectively to the coarsest and finest levels of numerical approximation.
While any of the approximations $\{\theta^0,\theta^1,\ldots,\theta^L\}$ can be used individually in Alg.~\ref{algo:SMC2} to approximate $Y$,
instead, the MLMC method independently estimates each of the expectations on
the right-hand side of \eqref{eq:linearity} such that the overall variance 
is minimized for a given computational cost.
The estimator $\hat{P}$ of $\E g\left(\theta^{L}\right)$ can be seen as a sum of independent estimators
\begin{equation}
\label{eq:mlmcest}
\hat{P}=\sum\limits_{\ell=0}^L P^{\ell},
\end{equation}
where $P^0$ is an estimator for
$\E g\left(\theta^{0}\right)$ based on $N_0$ samples, and $P^\ell$
are estimates for
$\E\left[ g\left(\tl\right) - g\left(\theta^{\ell-1}\right)\right]$
based on $N_\ell$ samples.
As we saw in the MSC method of Section \ref{sec:Monte_Carlo}, the simplest forms for $P^0$ and $P^\ell$ are the empirical means over all samples:
\begin{equation}
\label{eq:MLMC_empirical}
P^0 = \cfrac1{N_0}\sum\limits_{i=1}^{N_0} g\left(\theta^{0}_i\right),\quad
\ P^\ell = \cfrac{1}{N_\ell}\sum\limits_{i=1}^{N_{\ell}} \left[g\left(\tl_i\right) - g\left(\theta^{\ell-1}_i\right)\right],\,\quad\ell=1,\ldots,L.
\end{equation}
Using the assumption of having independent estimators $\{P^0,P^1,P^2,\ldots,P^L\}$ and employing the telescoping sum \eqref{eq:linearity} we can compute respectively the variance of $\hat{P}$ and bias as
$$\V\hat{P} = \V\left[\sum\limits_{\ell=0}^L P^{\ell}\right] = \sum\limits_{\ell=0}^L \V P^{\ell},\quad \E P - \E\hat{P} = \E P - \E\left[\sum\limits_{\ell=0}^L P^{\ell}\right] = \E P - \E g\left(\theta^{L}\right).$$

The computation of $P^\ell$ in \eqref{eq:MLMC_empirical} requires the samples $\theta^\ell_i,\theta^{\ell-1}_i$ to be generated from a common probability space.
We utilize the fact that sum of normal random variables is still normally distributed. 
Alg.~\ref{algo:dt-SHS:mlmc} presents generation of approximate coupled samples $\theta^\ell_i,\theta^{\ell-1}_i$ 
for the random variable $Y$ defined on the execution of a \SHS $\mathcal H$.
As can be seen in steps \ref{step:update_fine1}-\ref{step:update_fine2} and \ref{step:interp_couple1}, the approximate execution for the finer 
level $\ell$ is constructed exactly the same way as in Alg.~\ref{algo:dt-SHS} with $n_f = \kappa 2^\ell$ time steps.
The construction of approximate execution for the coarser 
level $(\ell-1)$ with $n_c = \kappa 2^{\ell-1}$ is also similar except that the noise term 
in step \ref{step:update_course} is obtained by taking the weighted sum of noise terms from the finer 
level $(W_{2k}^\ell+W_{2k+1}^\ell)/\sqrt{2}$. 
This choice preserves the properties of each approximation level while coupling the executions of
levels $\ell-1,\ell$ thus also coupling approximate samples $\theta^{\ell-1},\theta^\ell$.

\begin{algorithm}[t]
	\caption{Approximate coupled samples $\theta^\ell,\theta^{\ell-1}$ of random variable $Y$}
	\begin{center}
		\begin{algorithmic}[1]
			\label{algo:dt-SHS:mlmc}
			\REQUIRE 
			model $\mathcal H = \left(Q,\mathcal X, b, \sigma,x_0,r\right)$, $Y$ a function of execution of $\mathcal H$, time interval $[0,s]$, level $\ell$
			\STATE
			\label{step:ell_couple}
			select the number of time steps $n_f = \kappa 2^\ell$ and $n_c = \kappa 2^{\ell-1}$ for some $\kappa\ge 1$
			\STATE
			\label{step:Delta_couple}
			compute the time step $\Delta_c := s/n_c,\ \Delta_f :=s/n_f$ and set $k:=0$
			\STATE
			\label{step:init_couple}
			set the initial states $x_0^{\ell,c} : = (q_0^{\ell,c},z_0^{\ell,c})$ and $x_0^{\ell,f} : = (q_0^{\ell,f},z_0^{\ell,f})$ according to $x_0 = (q_0,z_0)\in X$
			
			\WHILE {$k<n_c$}
			\STATE
			sample $W_{2k}^\ell,\,W_{2k+1}^\ell$ independently from the standard $m$-dimensional normal distribution
			\STATE\label{step:update_fine1}
			update hybrid state $(q_{2k+1}^{\ell,f},z_{2k+1}^{\ell,f}) = \textsf{Update}(\mathcal H,q_{2k}^{\ell,f},z_{2k}^{\ell,f},\Delta_f,W_{2k}^\ell)$ using Alg.~\ref{algo:Euler}
			\STATE \label{step:update_fine2}
			update hybrid state $(q_{2k+2}^{\ell,f},z_{2k+2}^{\ell,f}) = \textsf{Update}(\mathcal H,q_{2k+1}^{\ell,f},z_{2k+1}^{\ell,f},\Delta_f,W_{2k+1}^\ell)$ using Alg.~\ref{algo:Euler}
			\STATE \label{step:update_course}
			update hybrid state $(q_{k+1}^{\ell,c},z_{k+1}^{\ell,c}) = \textsf{Update}(\mathcal H,q_{k}^{\ell,c},z_{k}^{\ell,c},\Delta_c,(W_{2k}^\ell+W_{2k+1}^\ell)/\sqrt{2})$ using Alg.~\ref{algo:Euler}
			\STATE $ k = k+1$
			\ENDWHILE
			\STATE
			\label{step:interp_couple1}
			define $z^{\ell,f}(t) = \sum\limits_{k=0}^{n_f}z_k^{\ell,f} 1_{[k\Delta_f,(k+1)\Delta_f)}(t)$ and $q^{\ell,f}(t) = \sum\limits_{k=0}^{n_f}q_k^{\ell,f} 1_{[k\Delta_f,(k+1)\Delta_f)}(t)$
			\STATE
			\label{step:interp_couple2}
			define $z^{\ell,c}(t) = \sum\limits_{k=0}^{n_c}z_k^{\ell,c} 1_{[k\Delta_c,(k+1)\Delta_c)}(t)$ and $q^{\ell,c}(t) = \sum\limits_{k=0}^{n_c}q_k^{\ell,c} 1_{[k\Delta_c,(k+1)\Delta_c)}(t)$
			\STATE
			\label{step:sample_couple}
			compute $\theta^\ell$ and $\theta^{\ell-1}$ by applying the definition of $Y$ to $(q^{\ell,f}(\cdot),z^{\ell,f}(\cdot))$ and $(q^{\ell,c}(\cdot),z^{\ell,c}(\cdot))$ respectively 
			\ENSURE
			$\theta^{\ell}$, $\theta^{\ell-1}$  as approximate sample of $Y$
		\end{algorithmic}
	\end{center}
\end{algorithm}

Now we are ready to present the MLMC method in Alg.~\ref{algo:MLMC_simplified}. The method is parameterized by the number of levels $L$, number of samples for each level $N_\ell$, $\ell= 0,1,\ldots, L$ (which are gathered in $\mathfrak S$), and the initial number of time steps $\kappa$. Steps~\ref{step:A_0}-\ref{step:P_0} performs the SMC method of Alg.~\ref{algo:SMC2} with embedded sampling algorithm~\ref{algo:dt-SHS} in order to estimate $\E g(\theta^0)$ with $N_0$ samples at the initial level $\ell=0$. Then the algorithm iteratively estimate $\E [g(\theta^l)-g(\theta^{l-1})]$ in steps \ref{step:A_l}-\ref{step:P_l} using Alg.~\ref{algo:SMC2} with number of samples $N=N_l$ and with the embedded coupled sampling algorithm~\ref{algo:dt-SHS:mlmc}.
The sum estimated quantity is reported in step \ref{step:hat_P} as the estimation of $\E g(Y)$.
\begin{algorithm}[ht]
	\caption{MLMC method to estimate $\E g(Y)$
	}
	\begin{center}
		\begin{algorithmic}[1]
			\label{algo:MLMC_simplified}
			\REQUIRE
			model $\mathcal H = \left(Q,\mathcal X, b, \sigma,x_0,r\right)$, $Y$ a function of execution of $\mathcal H$, time interval $[0,s]$, functional $g(Y)$
			\STATE select the parameters: finest level of approximation $L$, number of samples for each level $\mathfrak S := (N_0,N_1,\ldots,N_L)$, initial number of time steps $\kappa$
			\STATE \label{step:A_0}
			define $\A_0$ to be Alg.~\ref{algo:dt-SHS} with $\ell = 0$ and time step $n_0 = \kappa2^0$ that generates samples $\theta^{0}$ 
			\STATE \label{step:P_0}
			compute $P^0$ using Alg.~\ref{algo:SMC2} with number of samples $N=N_0$ and functional $g(\theta^0)$ and with the embedded algorithm~$\A_0$
			\STATE $l=1$
			\WHILE{$l<L$}
			\STATE \label{step:A_l}
			define $\A_l$ to be Alg.~\ref{algo:dt-SHS:mlmc} with time step $n_f = \kappa 2^\ell$ that generates samples $\theta^{\ell}$ and $\theta^{\ell-1}$
			\STATE \label{step:P_l}
			compute $P^\ell$ using Alg.~\ref{algo:SMC2} with number of samples $N=N_l$ and functional $[g(\theta^l)-g(\theta^{l-1})]$ and with the embedded algorithm~$\A_\ell$
			\STATE $l=l+1$
			\ENDWHILE
			\STATE \label{step:hat_P}
			compute $\hat P = \sum_{\ell=0}^L P^\ell$ according to \eqref{eq:mlmcest}
			\ENSURE
			$\hat P$ as approximate estimate of $\E g(Y)$
		\end{algorithmic}
	\end{center}
\end{algorithm}

The next theorem gives the convergence rate of MLMC method presented in Alg.~\ref{algo:MLMC_simplified}.
\begin{theorem}
	\label{mlmc:main}
	Let $\theta^\ell$ denote the level $\ell$ numerical approximation of the random variable $Y$.
	Assume the independent estimators $P_\ell$ used in Alg.~\ref{algo:MLMC_simplified} satisfy
	\begin{align}
	&\left| \E[g(\theta^\ell) \!-\! g(Y)] \right|\ \leq\ c_1\, 2^{-\alpha\, \ell}\quad\text{ and }\quad
	\mathbb{E}[C_\ell]\ \leq\ c_2\, 2^{\zeta\, \ell}\label{ass:MLMC1}\\
	& \mathbb{E}[P^\ell]\ = \left\{ \begin{array}{ll}
	\mathbb{E}[g(\theta^0)],                     &~~ \ell=0 \\[0.1in]
	\mathbb{E}[g(\theta^\ell) \!-\! g(\theta^{\ell-1})], &~~ \ell>0
	\end{array}\right.
	\quad\text{ and }\quad
	\V[P^\ell]\ \leq\ c_3\, N_\ell^{-1}\, 2^{-\beta\, \ell}\label{ass:MLMC2}
	\end{align}
	for positive constants $\alpha, \beta, \zeta, c_1, c_2, c_3$ with 
	$\alpha\!\geq\!{\textstyle \frac{1}{2}}\,\min(\beta,\zeta)$.
	Then the MLMC method in Alg.~\ref{algo:MLMC_simplified} converges with rate
	$2+\cfrac{\max(\zeta\!-\!\beta,0)}{\alpha}.$
\end{theorem}

Assumptions in \eqref{ass:MLMC1} are exactly the same as the ones used in Theorem \ref{thm:SMC}.
Assumptions in \eqref{ass:MLMC2} put restriction on the statistical properties of the estimators $P^\ell$:
they first enables us to use the telescoping property \eqref{eq:linearity} and the second ensures the exponentially decaying variance as a function of level $\ell$.  
In compare with the convergence rate of SMC method in Theorem \ref{thm:SMC}, the improvement is due to the non-zero factor $\beta$ which is the decaying rate of the variance of estimators.

\medskip

\noindent
\textbf{Application to the TCL Case Study.}
We construct the approximate discrete-time executions for the finer lever $\ell$ as 
\begin{align}
&\theta^{\ell,f}_{k+1} = \frac{1}{CR}(\theta_a - q^{\ell,f}_k R P_{rate}-\theta^{\ell,f}_k)\Delta_f+\sigma(q^{\ell,f}_k)\cdot\sqrt{\Delta_f}\cdot W^{\ell}_k,\label{euler:numer2}\\
&q^{\ell,f}_{k+1} := f(q^{\ell,f}_k,\theta_k^{\ell,f}),\qquad \text{ for all } k=0,1,\ldots,n_f,\nonumber
\end{align}
where $W^{\ell}_k$ is the sample from the standard normal distribution, $\Delta_f=s/n_f$, $n_f = \kappa 2^{\ell}$,
and with $f(\cdot)$ defined in \eqref{eq:switch}.
The coupling, which means that we get the dynamics for $\theta^{\ell,c}$ based on the increments for $\theta^{\ell,f}$, is done in a following way: 
\begin{align}
&\theta^{\ell,c}_{k+1} = \theta_k^{\ell,c}+\frac{1}{CR}(\theta_a- q^{\ell,c}_k R P_{rate}-\theta^{\ell,c}_k)\Delta_c + \sigma(q_k^{\ell,c})\cdot\sqrt{\Delta_c}\cdot\frac{1}{\sqrt{2}}\cdot(W^{\ell}_{2k-1}+W^{\ell}_{2k}),\label{euler:numer3}\\
&q^{\ell,c}_{k+1} := f(q^{\ell,c}_k,\theta_k^{\ell,c}),\qquad \text{ for all } k=0,1,\ldots,n_c,\nonumber
\end{align}
where $\Delta_c=s/n_c$ with $n_c = \kappa 2^{\ell-1}$.
The fact that we have used the same Brownian increments $W^{\ell}_{2k-1},W^{\ell}_{2k}$ from the finer level \eqref{euler:numer2} in the courser level \eqref{euler:numer3}
lays the foundation of having nonzero value of 
$\beta$ in Theorem \ref{mlmc:main}. The cost of simulating one approximate execution in \eqref{euler:numer2}-\eqref{euler:numer3} is proportional to the number of discretization steps, thus setting the parameter $\zeta=1$ in Theorems \ref{thm:SMC}-\ref{mlmc:main}.

\medskip

Now that we have set up the MLMC method and the coupling technique that improves the convergence rate of the estimation, 
we focus on the following important problems associated with the approach:
\begin{enumerate}
	\item Discontinuity of functional $g(Y)=1_{(-\infty,s]}(Y)$, leads to smaller values of $\alpha$ and $\beta$ in Theorem \ref{mlmc:main}.
	This results in larger convergence rate $\gamma$ thus larger computational cost for a given accuracy $\varepsilon$.
	\item The optimal choice of parameters $N_\ell$, $L$ and the unknown constants in Theorem \ref{mlmc:main}.
\end{enumerate}
The first issue, discussed in Section \ref{s2.1}, is resolved through \emph{smoothing}, 
which replaces the discontinuous function $g$ with a smoothed function $g^\delta$ with Lipschitz constant proportional to $\delta^{-1}$.
The second issue, discussed in Section~\ref{sec:algorithm}, is resolved through an adaptive algorithm. 
This adaptive algorithm follows \cite{giles2014multi}, and combines the smoothing of discontinuous functionals and the MLMC method.
Note that we require an updated set of assumptions and include the search for parameter $\delta$ into the adaptive algorithm.

\section{MLMC with Smoothed Indicator Function}
\label{s2.1}

The smoothing is based on the function $g^\delta:\R \to \R$, which are the rescaled translates of a function $g^0 : \R \to \R$  of the form
\begin{equation}
\label{func:sm}
g^0(x)=\begin{cases}
0, & x>1\\
\frac12+\frac18\left(5x^3 - 9x\right), &-1\le x\le1\\
1, &x<-1,
\end{cases}
\quad
\text{and}\quad
g^{\delta} (x) = g^0((x-s)/\delta),\ x \in \R.
\end{equation}
Since we add a smoothing step, we need to update the MLMC estimator \eqref{eq:mlmcest}, 
derive new a MSE decomposition (instead of \eqref{eq:mse}) 
which incorporates the error due to the smoothing, 
and update Assumptions \eqref{ass:MLMC1}-\eqref{ass:MLMC2} in Theorem \ref{mlmc:main}.

Note that function~\eqref{func:sm} is not the only possible choice for a smoothing function (see \cite{giles2014multi}), 
but in our experience this is the easiest to implement and numerically stable, while still providing significant gains 
in computational cost.

Recall that the MLMC method is based on
a sequence $(\theta^{\ell})_{\ell\in\N_0}$ of  
random variables, defined on a common probability space together with $Y$. The new MLMC method that includes smoothing
is defined by
\begin{equation}
\label{eq:MLMC_smoothed}
\MM
=
\frac{1}{N_{0}} \cdot \sum_{i=1}^{N_{0}} 
g^{\delta} (\theta^{0}_i) +
\sum_{\ell=1}^{L}
\frac{1}{N_\ell} \cdot \sum_{i=1}^{N_\ell} 
\left( g^{\delta} (\theta^{\ell,f}_i) - g^{\delta} (\theta^{\ell,c}_i)
\right),
\end{equation}
with an independent family of $\R^2$-valued random variables
$(\theta^{\ell,f}_i,\theta^{\ell,c}_i)$ for $i=1,\dots,N_\ell$ and $\ell=0,1,\dots,L$ 
such that equality in distribution holds for
$(\theta^{\ell,f}_i,\theta^{\ell,c}_i)$ and
$(\theta^{\ell},\theta^{\ell-1})$, where we used the notation $(\theta^{0,f}_i,\theta^{0,c}_i) = (\theta^{0}_i,0)$ for the initial level $\ell=0$.
Note that \eqref{eq:MLMC_smoothed} is the same as the MLMC estimator \eqref{eq:mlmcest} except using the smoothing function $g^\delta(\cdot)$ instead of the indicator function $g(\cdot)$.
The next theorem gives the mean square error decomposition for \eqref{eq:MLMC_smoothed}.
\begin{theorem}
	For $\delta > 0$, the error of $\MM$ in \eqref{eq:MLMC_smoothed} with smoothing function \eqref{func:sm} can be decomposed as
	\begin{align}
	MSE\left(\MM\right) &:= \E\|\MM-\E g(Y)\|^2 \notag\\
	&\leq
	\delta^4 + \bigl|\E (g^{\delta} (Y)) -\E (g^{\delta} (\theta^L))\bigr|^2+\V(\MM)
	=: e_1^2+e_2^2+e_3.
	\label{eqeqeq}
	\end{align}
\end{theorem}

The error terms in \eqref{eqeqeq} are related to smoothing, bias, and variance, respectively.
Note that as $\delta$ goes to zero, the Lipschitz constant for $g^\delta(x)$ goes to infinity, which has to be taken into account. 
Hence the assumptions in Theorem \ref{mlmc:main} have to be updated. 
The theoretical analysis and updated assumptions are presented in \cite{giles2014multi}.

\section{Adaptive MLMC Algorithm}\label{sec:algorithm}

In this section we present an adaptive algorithm to find the optimal parameters for the MLMC method.
For a given $\Eps >0$ we wish to select the parameters of the MLMC algorithm such that its error is at most $\Eps$ and its cost is as small as possible.
Our approach to the selection of the replication numbers and of the maximal level follows \cite{giles15}.

The adaptive algorithm assumes no prior knowledge on the smoothing parameter $\delta$, along with bias and variance dependencies on it.
The smoothing parameter $\delta$ is chosen from the discrete set of values $\delta_m = 1/ 2^{m},$ where $m \in \N$. 
With a slight abuse of notation we put $g^{m} = g^{\delta_m}.$
In order to achieve 
$MSE(\MM) \leq \Eps$
we have to assign certain proportions of $\Eps$ to the three sources of 
the error introduced in \eqref{eqeqeq}.
Specifically we wish
to choose the parameters of our algorithm such
that
\begin{equation}\label{eq222}
e_1
\le a_1\Eps_*, \quad
e_2
\le a_2 \cdot \Eps_*, \quad
e_3 \le a_3^2 \cdot \Eps_*^2,\text{ where } \Eps_* := \frac{\Eps}{a_1+a_2+a_3}.
\end{equation}
The MLMC algorithm is parameterized by the  value $m$ for smoothing $\delta_m = 1/2^m$, the values of the maximal level $L$, and the
replication numbers $\mathfrak S = \left(N_0,\ldots,N_L\right)$. We always select
$L \geq 2$ and $N_\ell \geq 100$ for $\ell=0,\dots,L$. 
By the latter, we ensure a reasonable accuracy in certain estimates
to be introduced below.
We use $y_{i,0}$ to denote \emph{actual samples} of the random variable $\theta^{0}$
and $(y_{i,\ell},y_{i,\ell-1})$ to denote the actual samples of the random vector
$(\theta^{\ell},\theta^{\ell-1})$ for $\ell=1,\dots,L$ as opposed to $\theta_i^{\ell,f},\theta_i^{\ell,c}$ which were used previously for their respective \emph{random variables}.

\smallskip
\noindent
\textbf{Assumptions.}
Theorem \ref{mlmc:main} relies on the assumption of exponential upper bounds in \eqref{ass:MLMC1}-\eqref{ass:MLMC2}, which in general might be difficult to verify. Instead in this section we study asymptotic upper bounds. For this purpose we use the following notation.
For sequences of real numbers $u_\ell$ and positive real numbers
$w_\ell$ we write $u_\ell \approx w_\ell$ if
$\lim_{\ell \to \infty} u_{\ell}/ w_{\ell} = 1,$
and write $u_\ell \ls w_\ell$ if
$\limsup_{\ell \to \infty} u_{\ell}/ w_{\ell} \leq 1.$ We also replace assumptions \eqref{ass:MLMC1}-\eqref{ass:MLMC2} with the requirement that
for every $m$ there exists $c,\alpha > 0$
such that
\begin{equation}\label{eq206}
|\E (g^{m} (\theta^{\ell})) -\E (g^{m} (\theta^{\ell-1})) |
\approx c \cdot 2^{-\ell\cdot\alpha}
\quad\text{ and }\quad
\lim_{\ell \to \infty} \E g^{m} (\theta^{\ell}) = 
\E g^{m} (Y).
\end{equation}
This yields the following asymptotic upper bound for the bias at level $\ell$
\begin{equation}\label{eq206a}
\bigl|\E (g^{m} (Y)) -\E (g^{m} (\theta^{\ell})) \bigr| \ls (2^\alpha-1)^{-1} \cdot
\bigl|\E (g^{m} (\theta^{\ell})) -\E (g^{m} (\theta^{\ell-1}))
\bigr|.
\end{equation}
We put $C_r = 2^{r+1}$ with $r=3$, the degree of polynomial in \eqref{func:sm}, and suppose that there exists $c >0$ such that
$\bigl|\E (g^{m} (Y)) -\E (g^{m-1} (Y))
\bigr|
\approx c \cdot \delta_m^4.$
This yields the asymptotic upper bound for the smoothing error with parameter~$\delta_m$, 
\begin{equation}\label{eq211a}
\bigl|\E g(Y) -\E (g^{m} (Y)) \bigr|
\ls (C_r-1)^{-1} \cdot
\bigl|\E (g^{m} (Y)) -\E (g^{m-1} (Y))\bigr|.
\end{equation}

Our adaptive MLMC algorithm is based on the intuition that the asymptotic bounds \eqref{eq206a} and \eqref{eq211a} can be replaced by their corresponding inequalities ($\le$ instead of $\ls$),
and estimators for means and variances can be assumed to be nearly exact.

\smallskip
\noindent
\textbf{Variance Estimation and Selection of the Replication Numbers.}
To estimate the expectations  and variances we employ the empirical mean and variance
\begin{align}
\label{eq205}
&\hat{b}_0 = \frac{1}{N_0} \cdot \sum_{i=1}^{N_0} 
g^{m}(y_{i,0}),\quad\text{ and }\quad\hat{b}_\ell = \frac{1}{N_\ell} \cdot \sum_{i=1}^{N_\ell} 
(g^{m}(y_{i,\ell}) - g^{m}(y_{i,\ell-1})),\\
\label{eq246}
&\hat{v}_0 = \frac{1}{N_0} \cdot \sum_{i=1}^{N_0} 
|g^{m}(y_{i,0}) - \hat{b}_0|^2\text{ and }
\hat{v}_\ell = \frac{1}{N_\ell} \cdot \sum_{i=1}^{N_\ell} 
|g^{m}(y_{i,\ell}) - g^{m}(y_{i,\ell-1})
- \hat{b}_\ell|^2.
\end{align}
We get that
$\hat{v}(\mathfrak S) =  
\sum_{\ell=0}^L \frac{1}{N_\ell} \cdot \hat{v}_\ell$
serves as an empirical upper bound for the variance of
the MLMC algorithm with any choice of replication numbers $\mathfrak S = \left(N_0,N_1,\ldots,N_L\right)$.
If, for the present choice of replication numbers, 
this bound is too large
compared to the upper bound for $\V(\MM)$ in \eqref{eq222},
i.e., if the variance constraint
\begin{equation}\label{eq203}
\hat{v}(\mathfrak S) 
\leq 
a_3^2 \cdot \Eps_*^2
\end{equation}
is violated,
we determine new values of $N'_0,\dots,N'_L$ by minimizing 
$c(N_0,\dots,N_L)$ subject to the constraint
$\hat{v}(\mathfrak S) \leq a_3^2 \cdot \Eps_*^2$, which leads to
\begin{equation}\label{eq247}
N'_\ell
= \cfrac{\hat{v}_\ell^{1/2}}{(2^{\ell} + 1)^{1/2}}
\cdot \sum_{\ell=0}^L 
\left( \hat{v}_\ell \cdot (2^{\ell} + 1)\right)^{1/2} \cdot 
\cfrac{\Eps_*^{-2}}{a_3^2},\qquad \ell = 0,1,\ldots, L,
\end{equation}
and extra samples of $\theta^{0}$ and
$(\theta^{\ell},\theta^{\ell-1})$ have to be generated accordingly. 

\smallskip
\noindent
\textbf{Bias Estimation and Selection of the Maximal Level.}
For estimating
$|\E (g^{m} (\theta^{\ell})) -\E
(g^{m} (\theta^{\ell-1})) |$
we can use the values of $|\hat{b}_\ell|$ already available from \eqref{eq205} for the levels $\ell=1,\dots,L$.
We estimate $\alpha$ and $c$ in \eqref{eq206} by a least-squares fit,
i.e., we take $\hat{\alpha}$ and $\hat{c}$ to minimize 
\begin{equation}
\label{eq:regression}
(\alpha,c) \mapsto
\sum_{\ell\in\cal{L}} 
\left( \log |\hat{b}_{\ell}| + 
\ell \cdot \alpha \, \log 2+ \log c \right)^2.
\end{equation}
While the value of $\hat{c}$ is irrelevant, 
an upper bound for
$\left|\E (g^{m} (\theta^{L})) -\E
(g^{m} (\theta^{L-1}))\right|$
is given by
$|\hat{b}_L|$,
or, more generally, by
$2^{(\ell-L) \cdot \hat{\alpha}} 
\cdot |\hat{b}_\ell|$ with $\ell \leq L$.
This geometric upper bound can be used to set the stopping criterion of increasing the maximal level. Let us define
\begin{gather}\label{eq238}
\hat{B}_2 =
\max \bigl(
|\hat{b}_2|,
|\hat{b}_{1}|/2^{\hat{\alpha}}
\bigr), \text{ for $L=2$} \\
\label{eq239}
\hat{B}_L =
\max \bigl(
|\hat{b}_L|,
|\hat{b}_{L-1}|/2^{\hat{\alpha}},
|\hat{b}_{L-2}|/2^{2 \hat{\alpha}}
\bigr)\text{ for $L \geq 3$.}
\end{gather}
The present value of $L$ is accepted as the maximal level, if 
the bias constraint
\begin{equation}
\label{bias:stop}
\hat{B}_L 
\leq 
a_2 \cdot (2^{\hat{\alpha}}-1) \cdot \Eps_*
\end{equation}
is satisfied.
Otherwise, $L$ is increased by one, and new 
samples will be generated. 

\smallskip
\noindent
\textbf{Selection of the Smoothing Parameter.}
We wish to determine the smallest value of $m$, i.e., the largest
value of $\delta_m$, such that 
\[
\bigl|\E (g^{m} (Y)) -\E (g^{m-1} (Y)) \bigr|
\leq 
a_1 \cdot (C_r-1) \cdot \Eps_*
\]
is satisfied, which corresponds to 
the upper bound for $e_1$ in \eqref{eq222} together with \eqref{eq211a}.
Initially we try $m=2$. 
Actually, $Y$ is approximated by $\theta^{L}$, so the present value is accepted if
\begin{equation}\label{eq234}
\hat{s} := 
\Bigl| \frac{1}{N_L} \cdot \sum_{i=1}^{N_L} 
(g^{m} (y_{i,L}) - g^{m-1} (y_{i,L}))\Bigr|\leq a_1 \cdot (C_r-1) \cdot \Eps_*.
\end{equation}

\smallskip
\noindent
\textbf{The Adaptive Algorithm.}
We combine the above results
and sum them up in Alg.~\ref{algo:MLMC:adaptive}, where the desired accuracy $\varepsilon$ is the input.

\begin{algorithm}[ht]
	\caption{Adaptive MLMC algorithm with smoothing}
	\begin{algorithmic}[1]
		\label{algo:MLMC:adaptive}
		\REQUIRE 
		sampling algorithm~$\A_\ell$, functional $g(\cdot)$, target accuracy $\varepsilon$
		\STATE initialize parameters $m=2$; $L=2$ $N_0=N_1=N_2=10^2$
		\STATE generate $N_0$ samples of $\theta^{0}$ and $N_\ell$ samples of 
		$(\theta^{\ell},\theta^{\ell-1})$ for $\ell=1,2$
		\STATE compute $\hat{v}_0,\hat{v}_1,\hat{v}_2$, according to \eqref{eq246}
		\REPEAT [/* smoothing */]
		\STATE $m = m+1$ and $\textsf{newlevel} = {\rm false}$
		\REPEAT[/* bias */]
		\IF {\textsf{newlevel}}
		\STATE $L = L+1$; $N_L = 100$
		\STATE generate $N_L$ number of samples of $(\theta^{L},\theta^{L-1})$
		\STATE compute $\hat{v}_L$ according to \eqref{eq246}
		\ENDIF
		\REPEAT
		\STATE  compute the replication numbers $N'_0,\dots,N'_L$ according to \eqref{eq247}
		\STATE $N_\ell = \max(N_\ell,N'_\ell)$ for $\ell=0,\ldots,L$
		\STATE generate extra samples of $\theta^{0}$ and 
		$(\theta^{\ell},\theta^{\ell-1})$ for $\ell=1,\ldots,L$
		\STATE compute $\hat{v}_0,\dots,\hat{v}_L$ according to \eqref{eq246}
		\UNTIL {the variance constraint \eqref{eq203} is satisfied}
		\STATE compute $\hat{\alpha}$ according to \eqref{eq:regression},
		and $\hat{B}_{L}$ according to \eqref{eq238}-\eqref{eq239}
		\STATE $\textsf{newlevel} = {\rm true}$
		\UNTIL {the bias constraint \eqref{bias:stop} is satisfied}
		\STATE compute $\hat{s}$ according to \eqref{eq234}
		\UNTIL {the smoothing constraint \eqref{eq234} is satisfied}
		\ENSURE $\MM$ as an estimation of $\E g(Y)$
	\end{algorithmic}
\end{algorithm}

\section{Simulation Results}
\label{sec:implementation}

Recall Problem \ref{specinter} where the goal is to estimate the probability $\mathbb{P}(Y\le \theta_+ + 0.1\cdot \delta_d)$. The random variable $Y$ is defined as $Y=\max\{\theta_t,\,t\in[0,s]\}$.
We set the parameters of the TCL model \eqref{eq:tcl_dyn}-\eqref{eq:switch} according to Table \ref{tab:het_parameters} and select the time horizon $s=1$~hour.
We implement the MLMC Alg.~\ref{algo:MLMC:adaptive} for target accuracies $\varepsilon=2^{-k}$, where $k\in\{3,\ldots,8\}.$
We set the parameters $a_1=4$, $a_2=a_3=2$ in \eqref{eq222}. With this choice we put less pressure on the smoothing error because the influence of the smoothing parameter $\delta$ on the variance and thus on the overal cost is severe.
Due to the smoothing step we have to sample executions for the time duration of at least $(s+\delta)$ in order to evaluate the functional $g(Y)$. With the selected values of $s$ and $\varepsilon$, sampling executions for $1.5$ hours is sufficient.

\begin{figure}[t]
	\centering
	\includegraphics[scale=0.38]{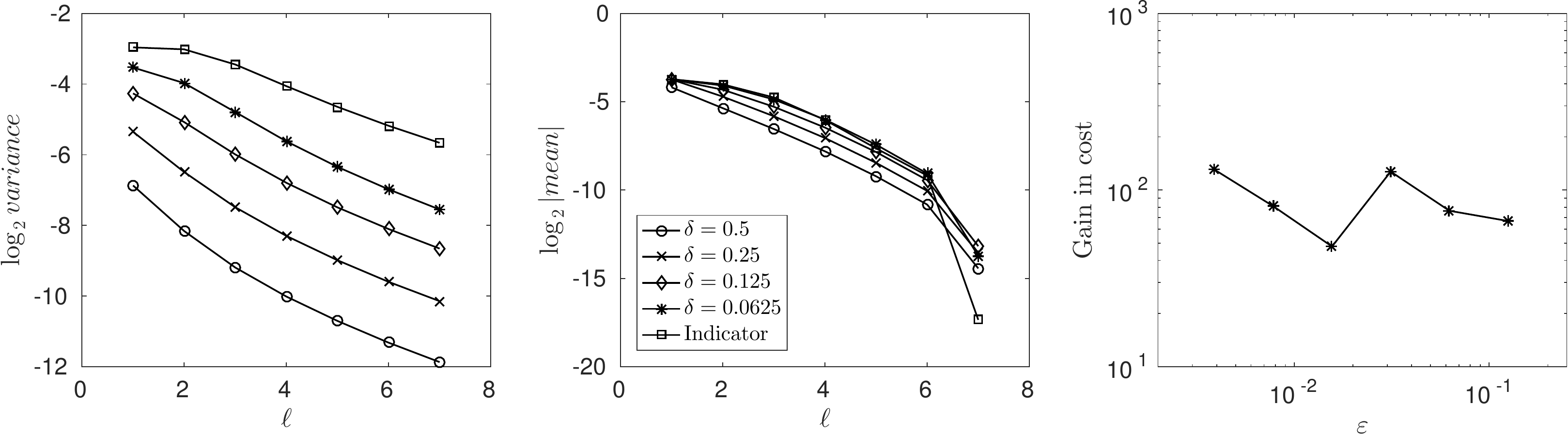}
	\caption{Simulation results for Problem \ref{specinter}. Variance (left) and mean (center) of the estimation decay with respect to level $\ell$ for different smoothing coefficient. Computational gain (right) is computed as ratio of the cost of SMC over cost of adaptive MLMC.}
	\label{fig:new}
\end{figure}

The result of the experiments is presented in Figure \ref{fig:new}.
The left and center plots show the impact of the smoothing coefficient on the variance and mean decays respectively based on $10^6$ runs of the algorithm.
The data points of the plots with $\ell=1$ and with the indicator function are related to the SMC method.
These plots indicate that the adaptive MLMC method is beneficial over SMC method due to the strong variance and mean decay with respect to level $\ell$ as well as the use of smoothing function instead of the indicator function.

The computational gain of the MLMC over SMC is presented on the right plot based on $100$~runs.
The plot compares the \emph{expected} cost of the SMC method with the \emph{estimated} cost of the adaptive MLMC method.
The cost of SMC method is given by $\varepsilon^{-2-\frac{1}{\bar\alpha}}$ (see Theorem \ref{thm:SMC}),
which bounds the cost of generating executions and evaluating functionals. We estimate the parameter $\bar\alpha$ through the precalculation and do not take into account the cost of estimating $\bar\alpha$. In this way we assume the parameter $\bar{\alpha}$ is known in advance and make the comparison more in favor of the SMC method.
The plot indicates larger computational gains for higher target accuracies (smaller $\varepsilon$). Note that the curve in the right plot is not monotone because there is an additional cost of updating the smoothing coefficient, hence re-evaluating the functionals with the new value of $\delta$. This additional cost has not been compensated by the MLMC gains as much in compare with the neighboring accuracies.

\begin{table}[t]
	\caption{Parameters of a residential air conditioner as a TCL \cite{SGEFA14} modeled in \eqref{eq:tcl_dyn}-\eqref{eq:switch}.}
	\begin{minipage}[b]{.41\textwidth}
		\centering
		\scalebox{0.92}[1]{
			\begin{tabular}{| l l l |}
				\hline
				Param. & Interpretation & Value\\
				\hline
				$\theta_s$ & set-point & $20\,[^\circ C]$\\
				$\delta_d$ & dead-band width & $0.5\,[^\circ C]$ \\
				$\theta_a$ & ambient temperature & $32\,[^\circ C]$ \\
				$P_{rate}$ & power & $14\,[kW]$\\
				\hline
			\end{tabular}
		}
	\end{minipage}
	\begin{minipage}[b]{.4\textwidth}
		\centering
		\scalebox{0.92}[1]{
			\begin{tabular}{| l l l |}
				\hline
				Param. & Interpretation & Value\\
				\hline
				$R$ & thermal resistance\hspace{0.1cm} & $1.5\,[^\circ C/kW]$ \\
				$C$ & thermal capacitance & $10\,[kWh/^\circ C]$\\
				$\sigma_0$ & standard deviation OFF mode & $0.2\,[^\circ C/\sqrt{hour}]$\\
				$\sigma_1$ & standard deviation ON mode & $0.22\,[^\circ C/\sqrt{hour}]$\\
				\hline
			\end{tabular}
		}
	\end{minipage}
	\label{tab:het_parameters}
\end{table}

\section{Conclusions}
\label{sec:concl}

In this paper we studied the problem of statistical model checking of continuous-time hybrid systems that do not admit exact simulations.
We employed multilevel Monte Carlo method and presented a smoothing step with tunable precision that replaces the desired discontinuous functional with a continuous approximation thus decreasing the overall computational effort of the approach.
An adaptive algorithm was designed which balances the errors due to the bias, variance, and smoothing.
The approach was demonstrated on the model of thermostatically controlled loads.



\bibliographystyle{plain}


\end{document}